\documentstyle[12pt]{article}
\def\J{$J/\psi$}
\def\j{J/\psi}
\def\P{$\psi'$}
\def\p{\psi'}
\def\U{$\Upsilon$}

\def\C{c{\bar c}}

\def\e{\epsilon}

\def\be{\begin{equation}}
\def\ee{\end{equation}}

\def\lsim{\raise0.3ex\hbox{$<$\kern-0.75em\raise-1.1ex\hbox{$\sim$}}}
\def\gsim{\raise0.3ex\hbox{$>$\kern-0.75em\raise-1.1ex\hbox{$\sim$}}}

\def\NP{{ Nucl.\ Phys.\ }}
\def\PL{{ Phys.\ Lett.\ }}
\def\PR{{ Phys.\ Rev.\ }}

\def\PRL{{ Phys.\ Rev.\ Lett.\ }}

\def\ZP{{ Z.\ Phys.\ }}

\begin{document}

\thispagestyle{empty}

\hfill BI-TP 97/22

\vfill

\centerline{\bf QCD \& QGP:}
\medskip
\centerline{\bf A SUMMARY$^{*)}$}
\vskip 0.8 truecm
\centerline{\bf Helmut Satz}
\medskip
\centerline{Fakult\"at f\"ur Physik, Universit\"at Bielefeld}
\par
\centerline{D-33501 Bielefeld, Germany}
\par
\centerline{and}
\par
\centerline{Theory Division, CERN}
\par
\centerline{CH-1211 Geneva 23, Switzerland}
\bigskip
\centerline{\bf Contents:}
\medskip
\leftskip 2.3 truecm
{1.\ The Thermodynamics of Quarks and Gluons}
\par
{2.\ Hard Probes: Colour Deconfinement}
\par
{3.\ Electromagnetic Probes: Chiral Symmetry Restoration}
\par
{4.\ Soft Probes: Equilibration and Expansion}
\par
{5.\ Conclusions}
\par
\leftskip 0 truecm

\bigskip

The aim of high energy nuclear collisions is to study strong
interaction thermodynamics in the the laboratory; we want to explore
colour deconfinement and the resulting new state of matter, the
quark-gluon plasma. Phenomenological models have done much to form the
concepts of the field, but today QCD provides the theoretical basis
for our understanding of hot and dense matter and for the tools
to probe it. I will therefore begin by summarizing recent results from
finite temperature lattice QCD and then turn to the study of colour
deconfinement using hard probes; here the recently reported anomalous
\J~suppression represents a particularly promising signal. Similarly,
the observed low mass dilepton enhancement has focussed our attention on
the properties of hadrons near chiral symmetry restoration. The
hadrosynthesis at freeze-out is yet another region of much present
activity, to be addressed in the final part of this summary.
\par
All aspects were covered here in a variety of excellent plenary talks
and contributions; I hope the speakers will forgive me for
concentrating on the progress in physics as I see it, rather than on
individual talks. The field of high energy nuclear collisions is very
many-faceted, and so I moreover had to select what I could
coherently summarize in the given time. I therefore also apologize to
all those whose contributions to this meeting are covered
insufficiently or not at all. In particular, I will review neither
developments in astrophysics nor the search for disoriented chiral
condensates, simply because of my lack of competence in these areas.

\bigskip

\hrule

\medskip

\noindent
$^{*)}$ Theory Summary, {\it International Conference on the
Physics and Astrophysics of\par\noindent~~ the Quark-Gluon Plasma
(ICPA-QGP'97)}, Jaipur/India, March 15 - 21, 1997.

\eject

\noindent
{\large{\bf 1.\ The Thermodynamics of Quarks and Gluons}}

\bigskip

Statistical QCD, as evaluated on the lattice \cite{Wilson} by means
of computer simulation \cite{Creutz}, is perhaps the only case in
statistical physics where critical behaviour can be calculated from
first principle dynamics \cite{Biele}, without having to invoke an
intermediate ``effective" theory. The precision of the predictions thus
obtained is limited only by computer performance -- with one rather
serious restriction: so far, the method is applicable (for what seem to
be technical reasons) only to matter at vanishing baryon density.
\par
The power of the appproach is best seen in the thermodynamics of pure
SU(2) and SU(3) gauge theory, i.e., for systems consisting of gluons
only. The evaluation of these is now essentially complete, largely
because of a fruitful interplay of finite-size scaling methods, improved
actions and the advent of more powerful computing facilities.
The main features considered are the deconfinement
transition and the properties of the hot gluon plasma. Below the
deconfinement temperature $T_c$, the constituents of the system are
colourless gluonium states (glueballs); above $T_c$, it consists of
coloured gluons. The transition was proposed to lie in the same
universality class as $Z_N$ spin systems of the same space dimension
\cite{Svet}; the critical exponents, which govern the singular behaviour
of the system at $T_c$, must then be the same for $SU(N)$ gauge and
$Z_N$
spin theory. We see in Table 1 that the SU(2) exponents, obtained for
the thermodynamic limit in a finite-size scaling analysis \cite{SU2},
indeed agree very well with those for the corresponding Ising model.
\par
$$
{\offinterlineskip \tabskip=0pt
\vbox{
\halign to 0.9\hsize
{\strut
\vrule width0.8pt\quad#
\tabskip=0pt plus100pt
& # \quad
&\vrule#&
&\quad \hfil # \quad
&\vrule#
&\quad \hfil # \quad
\tabskip=0pt
&\vrule width0.8pt#
\cr
\noalign{\hrule}\noalign{\hrule}
&~~~~~~~~~~~~~&&~~~~~~~~~~~~~~ &&~~          ~~~~~~~&\cr
& Exponent &&    SU(2)    &&     Ising       &\cr
&~~~~~~    &&~~~~~~~~~~~~~&&~~          ~~~~~&\cr
\noalign{\hrule}
%&~~        &&~~           &&                 &\cr
& $(1-\beta)/\nu$~
           &&  1.085(14)  &&  1.072(7)       &\cr
%&~~        &&~~           &&~                &\cr
\noalign{\hrule}
%&          &&             &&                 &\cr
& $(1+\gamma)/\nu$
           &&  3.555(15)  &&  3.560(11)      &\cr
%&~~        &&~~           &&~                &\cr
\noalign{\hrule}
%&          &&             &&                 &\cr
& $ \beta $
           &&  0.326(8)   &&  0.3258(44)     &\cr
%&~~        &&~~           &&~~               &\cr
\noalign{\hrule}
%&~~        &&~~           &&                 &\cr
& $ \gamma $
           &&  1.207(24)  &&  1.239(7)       &\cr
%&~~        &&~~           &&~~               &\cr
\noalign{\hrule}
%&          &&             &&                 &\cr
& $ \nu $
           &&  0.621(8)   &&  0.6289(8)      &\cr
%&~~        &&~~           &&~~               &\cr
\noalign{\hrule}}
}}
$$
\centerline{\bf Table 1:}

\medskip

\centerline{Critical exponents in SU(2) gauge theory and in the Ising
model \cite{SU2}}.

\bigskip

SU(3) gauge theory, just as the corresponding three-state Potts model,
leads to a first order transition, for which critical exponents are not
directly definable. Again, however, a finite-size scaling analysis can
be carried out to extrapolate to infinite spatial volume (thermodynamic
limit) and vanishing lattice spacing (continuum limit) \cite{SU3}. The
resulting equation of state, giving the energy density $\e$, the
pressure $P$ and the interaction measure $(\e-3P)$ as functions of
$T$, is shown in Fig.\ 1. Fixing the open dimension of the critical
temperature through the string tension $\sigma$, one obtains
\be
T_c/\sqrt{\sigma}= 0.629 \pm 0.003 , \label{1}
\ee
which leads to $T_c \simeq 260$ MeV for the quarkonium string tension
value $\sigma \simeq 420$ MeV. The latent heat of deconfinement is found
to be $\Delta \e /T_c^4 = 1.40 \pm 0.09$ \cite{Beinlich}. From the
temperature
behaviour of the interaction measure it is evident that in the region
$T_c \leq T \leq 2~T_c$ considerable plasma interactions remain as
possible remnants of confinement.
\par
We see in Fig.\ 1 that even at rather high temperatures ($T\simeq
5~T_c$), the thermodynamic variables are still some 10 - 15 \% below
the ideal gluon gas limit. One might expect such high temperature
deviations to be accountable by higher order perturbative corrections,
implying that we have reached a regime where lattice calculations
and perturbation theory meet. Recent calculations have provided
the perturbative corrections up to (the highest calculable) order $g^5$
\cite{perturb}; however, the results do not explain the deviations
from ideal gas behaviour found in lattice calculations. If the coupling
is chosen large enough (by suitably tuning the cut-off parameter)
to produce deviations of the observed size, the perturbation expansion
shows no signs of convergence; the $g^2$ to $g^5$ contributions
increase in magnitude and alternate in sign, so that their sum varies
strongly with the cut-off \cite{Karsch-Jai}. On the other hand, if
the cut-off in the
coupling is tuned to stabilize the perturbative result, then the overall
interaction effect is reduced to only 1 - 2 \%. The usefulness of
conventional perturbation theory in finite temperature gauge field
thermodynamics has thus become doubtful, creating the need for a
new method to treat the hot, near-ideal QGP.
\par
Such an approach may be given by ``screened" perturbation theory, in
which one starts from a gas of gluons having effective thermal masses.
It has been known for some time \cite{Karsch,Golo} that an ideal gas
of gluons of mass
\be
m_g(T) \simeq g(T)~T \label{2}
\ee
accounts well for the observed equation of state in SU(N) gauge field
thermodynamics, provided, however, that the massive gluons retain only
the two transverse degrees of freedom of their massless state. In a
recent study \cite{Patkos}, it was shown that such a scheme can be
formulated consistently in scalar $\phi^4$ theory, where conventional
perturbation theory encounters similar convergence problems. In Fig.\
2 we see that screened perturbation theory here leads to much better
convergence.
\par
For full QCD, i.e., for SU(3) gauge theory with dynamical quarks,
there exist today a number of studies on finite lattices, but not enough
yet to carry out a finite-size scaling analysis as in pure gauge theory.
To arrive also here at the thermodynamic and continuum limits, more
powerful computers are needed, and these are expected to become
available within the next few years.
\par
So far, we have some idea of the equation of state in the presence of
$N_f$ massless quarks \cite{MILC,Laermann}.
We know that chiral symmetry restoration
coincides with deconfinement, and that for presently used bare quark
masses (still too large for a correct pion mass), the sharpness of the
two transitions is quite comparable \cite{Laermann}, as seen in Fig.\
3. It remains unclear at this time \cite{Karsch-Jai,Kanaya-Jai} if
vanishing quark masses lead to a chiral transition in the same
universality class as the O(4) spin model \cite{Wilczek}. A
particularly interesting recent study indicates that the case
of real physical interest, two light ($u$ and $d$) and one heavier ($s$)
quark, may well lead to a first order transition
\cite{Kanaya-Jai,Kanaya}. Hopefully this will soon be clarified
by further and more detailed work. In addition, the study of quarkonium
states at finite temperature is under way. The most crucial
open challenge for lattice gauge theory thus remains the case of
non-vanishing baryon density, for which, as already mentioned, so far
no viable approach exists.
\par
To summarize the status of the thermodynamics of quarks and gluons:
statistical QCD predicts the hadron-quark deconfinement transition and
the properties of the QGP through first principle calculations of ever
increasing precision.
\par
We can thus turn to the task of studying these phenomena in high energy
nuclear collisions. Here the environment will be a rapidly evolving
medium, and we will have to find probes for the evolution stages of
interest. In this connection, let me quote a bit of advice which guide
books give travellers to many countries, including India. They say that
if you're lost, you should never ask a local farmer ``does this road
go to Jaipur?" Out of politeness, he will always agree, even if the road
in fact gets you nowhere near Jaipur; instead, you should ask where the
road in question leads. Our probes of nuclear collisions are probably at
least as polite, and so you should never ask them if they found the QGP;
it appears they'll always say that they did. Instead, we should just
ask them what they did see.
\bigskip

\noindent
{\large{\bf 2.\ Hard Probes: Colour Deconfinement}}

\bigskip

The basis for the existence of a deconfined state of matter is that a
high density of colour charges leads to a screening of long range
confining forces, so that only short range interactions remain
operative. How can we probe full or partial colour deconfinement in
systems produced in nuclear collisions? Any deconfinement probe
\begin{itemize}
\item must be present in the {\it early} stages of the collision
evolution,
\item must be {\it hard} enough to resolve sub-hadronic scales,
\item must be able to {\it distinguish} confinement and
deconfinement, and
\item must {\it retain} the information throughout the collision
evolution.
\end{itemize}
The latter feature implies that the probe should not be in thermal
equilibrium with the later evolution stages, since this would lead to a
loss of memory of previous stages. In the following we shall consider
two candidates for probing colour deconfinement: quarkonium states,
which show different dissociation patterns, and hard jets, which
suffer different energy losses in confined and in deconfined media.
\par
Quarkonium ground states (\J, \U) are small and tightly bound resonances
of heavy quarks. The \J, which we shall consider as prototype, has a
radius of about 0.2 fm, much smaller than the normal hadronic scale
$\Lambda_{\rm QCD}^{-1} \simeq 1$ fm; its binding energy is with 0.6
GeV much larger than $\Lambda_{\rm QCD} \simeq 0.2$ GeV. It therefore
requires hard gluons to resolve and dissociate a \J. As a consequence,
the collision of a \J~with conventional ``light" hadron probes the
gluon sub-structure of the light hadron, not its size or mass.
\par
From deep inelastic scattering experiments, the gluon distribution in
a light hadron is found to be
\be
x g(x) \simeq x^{-a} (1-x)^b, \label{3}
\ee
where $a,b$ are constants determined by experiment and/or quark
counting rules. The effect of such a
distribution on \J-hadron interactions is illustrated in Fig.\ 4 for
the case of elastic forward \J~photoproduction \cite{KSSZ}. Compared to
normal hadronic
behaviour, Eq.\ (3) leads to a strong threshold suppression caused by
the factor $(1-x)^b$, with $b>0$: slow hadrons relative to the \J~do
not contain sufficiently hard gluons to resolve the quark structure of
the small \J. On the other hand, at high energies, the observed
anomalous small $x$ behaviour \cite{HERA} of the proton structure
function implies $a>0$ and leads to an increasing cross section
(requiring normal hadron phenome\-no\-lo\-gy to introduce an additional
``hard" Pomeron). -- A further case where the large $x$ behaviour of the
gluon distribution becomes crucial was recently observed in the
reaction $\p \to \j~ \pi^+ \pi^-$ \cite{psi'}.
\par
And it is again this damping of the gluon distribution at large $x$
which makes the inelastic \J-hadron cross section become negligibly
small for hadron momenta below 3 - 5 GeV relative to a \J~at
rest \cite{KS3}. In contrast, the inelastic cross section for incident
deconfined gluons peaks around 1 GeV, corresponding to the photo-effect
in QCD. A comparison of the $g-\j$ and $h-\j$ cross sections as
functions of the respective projectile momentum $k$ is shown in Fig.\ 5.
Since $k \simeq 3T$, we conclude that confined matter for temperatures
up to about 600 MeV cannot dissociate a \J, whereas deconfined matter
for $T \geq 200$ MeV easily can. \J~suppression thus provides an
unambiguous test for colour deconfinement.
\par
Here two caveats should be added. The basic theoretical input, the
strong threshold suppression of the inelastic $\j-h$ cross section,
is corroborated by the mentioned experiments on \J~photoproduction and
on \P~decay. It can and should, however, also be tested directly in the
so-called ``inverse kinematics" experiment \cite{KS5}. -- The charmonium
test for colour deconfinement as discussed here applies to physical
\J's. In nuclear collisions, there will in addition be pre-resonance
nuclear absorption, and this has to be taken into account properly,
using information from p-A data \cite{KLNS}.
\par
Once that is done, these considerations can be applied to \J~data. It
appears \cite{Gonin}-\cite{Gonin-Jai} that nuclear collisions up to
central S-U interactions show only ``normal" pre-resonance absorption
in nuclear matter; in contrast, Pb-Pb collisions lead to a further strong
``anomalous" \J~suppression (Fig.\ 6), which can be interpreted as the
onset of colour deconfinement, though not necessarily in an equilibrated
medium \cite{K-Jai,B-Jai}. Such a conclusion would be supported by the
observation of an anomalous $p_T$-behaviour of \J~suppression in Pb-Pb
interactions, as recently predicted \cite{KNS}. Moreover, the crucial
feature of the observed effect, its sudden onset between S-U and Pb-Pb
collisions, must certainly be checked by experiments using different A-A
combinations and different incident energies.
\par
The use of hard jets as deconfinement probe has a similar basis. Hard
partons are formed at very early times, similar to the $\C$ formation
for charmonium. If such a parton travels through a deconfined medium, it
finds much harder gluons to interact with than it would in a confined
medium, where the gluons are constrained by the hadronic parton
distribution Eq.\ (3). As a result, jets will suffer a much
greater energy loss per unit length in a QGP than in hadronic
matter \cite{BDMPS}. It should be underlined, however, that to use
jet suppression as probe for the confinement status of a given medium,
we must know the ``normal" suppression in nuclear matter. Hence p-A
experiments will also here be essential.
\par
In closing this section, we note that charmonium states and jets
probe colour deconfinement in a rather general way. In both cases,
suppression does not imply an equilibrated medium, but rather a medium
containing gluons which are no longer subject to hadronic
parton distribution functions. Thus the smallest region of deconfinement
could arise in the overlap of two distinct nucleon-nucleon collisions.
If much of the medium were of this nature, we would speak of a QGP.
\bigskip

\noindent
{\large{\bf 3.\ Electromagnetic Probes: Chiral Symmetry Restoration}}

\bigskip

Real or virtual photons emitted during the evolution of the collision
subsequently undergo no (strong) interactions with the medium and hence
reflect its state at the time they were produced. On the other hand,
they are emitted during the entire collision evolution, and by
different dynamical mechanisms at different stages:
\begin{itemize}
\item Early hard parton interactions produce hard photons and
Drell-Yan dileptons; these provide information about the initial
(primary or pre-equilibrium) stages.
\item Thermal photon and dilepton emission by the medium,
through quark or hadron interactions, occur through its entire
evolution, and hence give information about the successive stages,
from QGP to final hadronic freeze-out.
\item Hadrons produced at any point of the hadronic stage, from
the quark-hadron transition to freeze-out, can decay and thereby
emit photons or dileptons; depending on the hadron decay time, they
provide information about dense interacting hadronic matter or about
hadrosynthesis at the end of the strong interaction era.
\end{itemize}
Let us consider each mechanism and its use as probe in some more detail.
\par
Drell-Yan dileptons and hard photons are the tools to study the
effective initial state parton distributions; in particular, they will
show any nuclear modifications (shadowing, anti-shadowing, coherence
effects) of these distributions. They also indicate the initial state
energy loss and the initial state $p_T$ broadening suffered by partons
in normal nuclear matter. Since they do not undergo any final state
strong interactions, they moreover provide a reference for
the effect of the produced medium on quarkonium states or jets. --
It should be noted that if measurable, open charm or beauty production
would give complementary information concerning these aspects
\cite{Sridhar}.
\par
Thermal emission can in principle serve as a thermometer for the
different evolution stages. The {\it functional form} of thermal
spectra,
\be
dN/dk_{\gamma} \sim e^{-k_{\gamma}/T} \label{4}
\ee
for photon momenta, or the corresponding distributions in the
dilepton mass $M_{l^+l^-}$, indicate the temperature $T$ of the medium
at the time the signal was emitted. The crucial problem here is to find
a ``thermal window", since the measured spectra are dominated at high
photon momenta or dilepton masses by hard primary reactions and at low
momenta or masses by hadron decay products. So far, thermal
photon or dilepton emission has appearently not been observed, perhaps
because of the dominance of the hadronic stage at present energies.
The situation may well become more favorable at RHIC or LHC, where one
may expect longer life-times for the hot medium.
\par
Since the functional form (4) for thermal production is the same
for a hadronic medium and for a QGP, it cannot specify the nature of the
emitting medium. There were attempts, however, to use the {\it rate} of
thermal emission as an indirect probe for the composition of the system.
But because of the evolution of the medium, it seems that this is a
rather model-dependent procedure; moreover, it was shown recently
\cite{Sriv,Sriv-Jai} that a purely hadronic resonance gas and an
evolving QGP with mixed phase and subsequent hadronisation can lead to
very similar results.
\par
The dileptons produced by hadron decay constitute an ideal tool to probe
in-medium hadron modifications, provided the hadrons actually decay
within the me\-dium. The $\rho$, with a half-life of about a fermi,
appears to be
the best candidate for such studies. Chiral symmetry restoration is
expected to change the properties of hadrons as the temperature of the
medium approaches the restoration point \cite{chiral}; hence such
in-medium changes are of particular interest, since they might be the
only experimental tool to address the chiral aspects of deconfinement.
\par
The low mass dilepton enhancement observed by the HELIOS and
CERES collaborations at CERN provides the experimental basis for
such studies \cite{Ullrich}-\cite{D-Jai}. In S-Au and Pb-Au
collisions, one finds in the mass region below the $\rho$ (from about
200 to 600 MeV) considerably more dilepton production than expected
from known hadronic sources ; these do provide the measured
distribution in p-A collisions. Thus some new effect appears to set in
as we go to nucleus-nucleus collsions.
\par
If at the onset of chiral symmetery restoration, the mass of the $\rho$
decreases sufficiently much \cite{Brown},
\be
{m_{\rho}(T)\over m_{\rho}(T=0)}~ \to~ 0~~{\rm as} ~~T~\to~T_c,
\label{5}
\ee
then the observed effect can be accounted for (Fig.\ 7). We note,
however, that the required drop of the in-medium $\rho$ mass (some
50~\%) is considerably larger than anything so far observed in finite
temperature lattice QCD. On the other hand, the meson masses studied
there are generally obtained from correlations and thus correspond to
screening masses rather than to those of physical states; hence
considerable uncertainties remain also on the level of lattice QCD.
\par
A recent alternative account is based on interaction broadening and
leads to in-medium changes of resonance {\it widths}, rather than {\it
masses} \cite{Friman,Wambach}.
A very much broader $\rho$, with the applicable kinematic constraints,
is found to also produce something like a low-mass dilepton enhancement
(Fig.\ 8). One would obviously like to find some distinguishing
feature for mass vs. width changes; one possible candidate is the $p_T$
dependence of the enhancement \cite{Friman,Wambach}.
\par
Whatever the underlying mechanism is eventually found to be, the low
mass dilepton enhancement seems to indicate the production of a dense
interacting ha\-dro\-nic medium in nuclear collisions.
\bigskip

\noindent
{\large{\bf 4.\ Soft Probes: Equilibrium and Expansion}}

\bigskip

The spectra and relative abundances of the usual ``light" hadrons
produced in nuclear collisons provide direct information on the state of
the system at the end of the strong interaction era. Two aspects have
here recently attracted particular attention.
\par
If the system at freeze-out is a gas of hadrons in full (thermal and
chemical) equilibrium, its temperature $T$ and baryo-chemical potential
$\mu$ determine the relative abundances of the emitted hadron species.
All measured production ratios of ground state hadrons and hadron
resonances (this can be 20 - 30 ratios!) would thus be given in terms
of only two parameters -- a very stringent test of equilibration,
which, if affirmative, would provide an unambiguous way to determine
the thermal freeze-out conditions.
\par
This test was recently applied to $e^+e^-,~pp$ and $p{\bar p}$
collisions \cite{Becat,B-H}, with the surprising result that even
these
elementary reactions lead {\it almost} to equilibrium ratios. Only the
(long known) suppression of strange particle production causes some
deviations, and after the introduction of partial strangeness saturation
$\gamma_s$ as single further parameter in addition to $T$ and $\mu$, one
obtains a remarkably good description of up to 30 different ratios
(Fig.\ 9), with $T\simeq 170$ MeV and $\gamma_s \simeq 0.5$.
 This ``Hagedorn-Becattini enigma" -- why do elementary
interactions lead to thermal hadron abundances? -- illustrates once more
that an equilibrium system has lost the memory of its formation. An
equilibrium hadron gas could be produced through sufficient multiple
scattering of primary and secondary hadrons; yet it is difficult to
imagine that this has happened in the restricted space-time extension
of $e^+e^-$ or $pp$ collisions.
\par
For A-A collisions, the main questions concerning particle ratios thus
are if the thermal composition persists and if the increase in
nucleon-nucleon interactions and space-time volume leads to an
equilibration  between the strange and the non-strange sectors
\cite{St,Cley-S}. If we would here encounter also full chemical
equilibration ($\gamma_s \to 1$), then we could indeed conclude that in
nuclear collisions ``there is nothing strange about strangeness"
\cite{Cleymans}. A first look at SPS data (Fig.\ 10) does show thermal
behaviour with essentially the same $T$ and some increase of
$\gamma_s$ \cite{PBM}-\cite{Sollfrank}, but a conclusive answer
to this question will probably have to wait until we have a sufficient
number of ratio measurements from Au-Au and Pb-Pb collisions.
\par
Changes in the functional form of hadronic spectra should probe the
presence of collective effects in the medium. For some years, this
question was mainly considered in terms of transverse hydrodynamic
flow, which had been shown to result in a mass-dependent broadening of
transverse momentum distributions \cite{Heinz?}. In the past year,
however, renewed attention was drawn to the fact that there is a
``normal" $p_T$-broadening observed in all reactions involving nuclear
targets, from Drell-Yan dilepton production to that of low $p_T$
mesons or baryons. For high $p_T$ hadrons, this is generally
referred to as Cronin effect \cite{Cronin}. Here as well as in
Drell-Yan or quarkonium production, it is accounted for by the fact
that successive parton scatterings rotate the collision axis relative
to the beam axis: any given transverse momentum distribution will
appear broadened when it is measured in the reference frame fixed by
the incident primary beams.
\par
These considerations were recently applied to low $p_T$ hadron
production in nuclear collisions \cite{Leonidov}, assuming that
successive collisions in nuclear reactions lead to a random walk in the
transverse momentum plane. The displacement $\delta$ per collision in
transverse rapidity was determined from p-A interactions;
the normalized $p_T$ spectra for A-B collisions are then predicted
parameter-free and found to agree quite well with preliminary data
from S-W \cite{S-W} and Pb-Pb \cite{Pb-Pb} interactions (Fig.\ 11). In
particular,
this ``normal" $p_T$-broadening also reproduces the increase with
increasing hadron mass, with more broadening for nucleons than for
kaons, and more for kaons than for pions. A very recent study
\cite{Kapu-Jai,Kapusta} has gone even further and determined the
``kick per collision" from p-p rather than from p-A data.
\par
At present, we thus find that the $p_T$-broadening observed in nuclear
reaction can be quite well accounted for by random walk collision axis
rotations, apart from possible resonance deviations at very small $p_T$.
Perhaps one might wish to consider such a phenomenon as a precursor
of ``transverse flow". Nevertheless, any hydrodynamic description of
$p_T$-distributions from A-B collisions, with the flow velocity as
open parameter, has to face two
tantalizing questions: why is there also broadening in p-A
interactions? and why can the ``flow velocity" in a random walk
approach be determined from p-A or even p-p interactions?
Perhaps only two-particle correlations, rather than single particle
spectra, can distinguish between hydrodynamic flow and a random walk
approach \cite{Heinz}.
\par
We close this section with some conceptual remarks. If it is indeed
observed experimentally that
\begin{itemize}
\item the relative abundances of the most copiously produced hadrons are
those of a thermal resonance gas, and
\item the $p_T$ broadening of their spectra follows a random walk
pattern,
\end{itemize}
then these are empirical features calling for an explanation. However,
the ``obvious" one, through multiple scattering of primary and
secondary hadrons, is phy\-si\-cally not tenable. The incident proton in
a
high energy p-A collision cannot execute a random walk through a target
nucleus which it passes in a proper time of less than 0.1 fm, just as in
a p-p collision the produced hadrons cannot scatter until they
form an equilibrium hadron gas. Hence finding the real origin of such
features is left as home-work for all of us; their occurrence also in
Drell-Yan production and $e^+e^-$ collisions may well be a hint to
look for the solution on a partonic level.

\bigskip

\noindent
{\large{\bf 5.\ Conclusions}}

\bigskip

QCD thermodynamics, as evaluated by computer simulation on the lattice,   leads to the quark-hadron transition and the existence of the
QGP as a new state of matter; calculations are so far performed at
vanishing overall baryon density. At a temperature of about 150 - 200
MeV, deconfinement sets in and chiral symmetry is restored; latest
lattice studies give some indication that for the three-flavour case,
with light $u/d$ quarks and a heavier $s$ quark, the transition may
well be of first order.
\par
Hard probes provide a direct test of colour deconfinement in nuclear
collisions. The dissociation pattern of quarkonia and the energy
loss of hard jets depend on the confinement status of the medium in
question: in both cases, this is based essentially on the hardening of
gluons no longer confined to hadrons. Since \J~dissociation requires
hard gluons, the anomalous \J~suppression recently observed in Pb-Pb
collisions could be a first indication of deconfinement.
\par
Electromagnetic probes, in particular low mass dileptons, can be used as
direct test for in-medium changes of hadron properties. Such
modifications are expected at the onset of chiral symmetry restoration
and thus constitute a way to address this aspect of the quark-hadron
transition. The presently observed low-mass dilepton enhancement could
be a first instance of such an effect for the $\rho$; however, an
alternative explanation through interaction broadening of the
$\rho$-width so far remains also tenable.
\par
Soft probes test equilibration and the presence of collective effects
at freeze-out, i.e., at the end of the strong interaction era for the
little bang. Recent studies of hadron abundances are in quite good
agreement with a composition as given by a thermal resonance gas; so
far, however, some strangeness suppression still remains. The
broadening of transverse momentum spectra observed in p-A and A-B
collisions agree well with random walk rotations of the collision axis.
Species equilibration as well as $p_T$-broadening in nuclear collisions
do not seem understandable in simple hadronic terms; however, a
consistent partonic description is also still lacking.

\bigskip\bigskip

\noindent
\centerline{\large{\bf Acknowledgements}}

\bigskip

It is a pleasure to thank the organizers of this conference for
providing such a stimu\-lating meeting in a part of the world
which through the ages has seen many intellectual
challenges. In preparing this talk, I have benefitted much
from discussions, agreements and disagreements with many collegues, in
particular with R.\ Baier, {J.-P.}\ Blaizot, P.\ Braun-Munzinger, J.\
Cleymans, U.\ Heinz, F.\ Karsch, D.\ Kharzeev, C.\ Louren{\c c}o, L.\
D.\ McLerran, M.\ Nardi, K.\ Redlich, J.\ Schukraft, J.\ Sollfrank, H.\
J.\ Specht, D.\ K.\ Srivastava, J.\ Stachel, R.\ Stock and H.\
Str\"obele; I am very grateful to all of them.

\vfill

\bigskip

\begin{thebibliography}{99}
\bigskip
\bibitem{Wilson}K.\ G.\ Wilson, \PR D 10 (1974) 2445.

\bibitem{Creutz}M.\ Creutz, \PR D 21 (1980) 2308.

\bibitem{Biele}
L.\ D.\ McLerran and B.\ Svetitsky, \PL 98 B (1981) 195; \par
J.\ Kuti, J.\ Pol\'onyi and K.\ Szlach\'anyi, \PL 98B (1981) 199; \par
J.\ Engels et al., \PL 101 B (1981) 89.

\bibitem{Svet}B. Svetitsky and L. Yaffe, \NP B 210 [FS6]
(1982) 423.

\bibitem{SU2}J. Engels et al., \PL B 365 (1996) 219.

\bibitem{SU3}G. Boyd et al., \NP B 469 (1996) 419.

\bibitem{Beinlich} B. Beinlich et al., \PL B 390 (1997) 268.

\bibitem{perturb}P. Arnold and C. Zhai, \PR D 50 (1994) 7603,
D51 (1995) 1906; \par
A. Nieto, hep-ph/9612291.

\bibitem{Karsch-Jai}F. Karsch, talk at this meeting.

\bibitem{Karsch}F. Karsch, \NP B 9 [Proc. Suppl.] (1989) 357.

\bibitem{Golo}V. Goloviznin and H. Satz, \ZP C 57 (1993) 671.

\bibitem{Patkos}F. Karsch, A.Patkos and P. Petreczky,
hep-ph/9702376, Feb. 1997.

\bibitem{MILC}L. K\"arkk\"ainen et al., \NP B (Proc. Suppl.) 42
(1995) 460.

\bibitem{Laermann}F.\ Karsch and E.\ Laermann, \PR D 50 (1994)
6954.

\bibitem{Kanaya-Jai}K. Kanaya, talk at this meeting.

\bibitem{Wilczek}R. Pisarski and F. Wilczek, \PR D 29 (1984)
339.

\bibitem{Kanaya}Y. Iwasaki et al., \PR D 54 (1996) 7010.

\bibitem{KSSZ} D.\ Kharzeev, H.\ Satz, A.\ Syamtomov and G.\ Zinovjev,
``\J~photoproduction and the parton structure of the nucleon", Bielefeld
preprint, in preparation.

\bibitem{HERA}M.\ Derrick et al. (ZEUS), \ZP C 72 (1996) 399;\par
S. Aid et al. (H1), \NP B 470 (1996) 3.

\bibitem{psi'} T.\ A. Armstrong et al., \PR D 55 (1997) 1153.

\bibitem{KS3}D.\ Kharzeev and H.\ Satz, \PL B 334 (1994) 155.

\bibitem{KS5}D.\ Kharzeev and H.\ Satz, \PL B 356 (1995) 365.

\bibitem{KLNS}D. Kharzeev et al., \ZP C 74 (1997) 307.

\bibitem{Gonin}M.\ Gonin [NA50], \NP A 610 (1996) 404c.

\bibitem{Carlos}C.\ Louren{\c c}o [NA50], \NP A 610 (1996) 552c.

\bibitem{Gonin-Jai}M.\ Gonin [NA50], talk at this meeting.

\bibitem{K-Jai}D.\ Kharzeev, talk at this meeting.

\bibitem{B-Jai}J.-P.\ Blaizot, talk at this meeting.

\bibitem{KNS}D.\ Kharzeev, M.\ Nardi and H.\ Satz,
hep-ph/9702273; \PL B, in press.

\bibitem{BDMPS}R.\ Baier et al., \NP B 484 (1997) 265.

\bibitem{Sridhar}H. Satz and K. Sridhar, \PR D 50 (1994) 351.

\bibitem{Sriv}J.\ Cleymans, K.\ Redlich and D.\ K.\ Srivastava,
\PR C 55 (1997) 1431.

\bibitem{Sriv-Jai}D.\ K.\ Srivastava, talk at this meeting.

\bibitem{chiral}R.\ Pisarski, \PL B 110 (1982) 155; \par
G.\ E.\ Brown and M Rho, \PRL 66 (1991) 2720; \par
T.\ Hatsuda and S.-H.\ Lee, \PR C 46 (1992) 34.

\bibitem{Ullrich}T.\ Ullrich [CERES], \NP A 610 (1996) 317c.

\bibitem{Drees}A.\ Drees [CERES], \NP A 610 (1996) 536c.

\bibitem{D-Jai}A.\ Drees [CERES], talk at this meeting.

\bibitem{Brown}G.\ Q.\ Li, C.\ M.\ Ko and G.\ E.\ Brown, \PRL
75 (1995)4007.

\bibitem{Friman}B.\ Friman and H.\ J.\ Pirner,
nucl-th/9701016, January 1997.

\bibitem{Wambach}R.\ Rapp, G.\ Chanfray and J.\ Wambach,
hep-ph/9702210, January 1997.

\bibitem{Becat}F.\ Becattini, \ZP C 69 (1996) 485.

\bibitem{B-H}F.\ Becattini and U.\ Heinz, hep-ph/9702274; \ZP C in
press.

\bibitem{St}P.\ Braun-Munziger et al., \PL B 344 (1995) 43.

\bibitem{Cley-S}J.\ Cleymans et al., \ZP C 74 (1997) 319.

\bibitem{Cleymans}J.\ Cleymans, talk at this meeting.

\bibitem{PBM}P.\ Braun-Munziger et al., \PL B 365 (1996) 1.

\bibitem{PBM-Jai}P.\ Braun-Munziger, talk at this meeting.

\bibitem{St-SPS}J.\ Stachel, \NP A 610 (1996) 509c.

\bibitem{Sollfrank}F.\ Becattini, J.\ Sollfrank and H.\ Str\"obele,
private communication.

\bibitem{Heinz?}K.\ S.\ Lee and U.\ Heinz, \ZP C 43 (1989) 425.

\bibitem{Cronin}J.\ W. Cronin, \PR D 11 (1975) 3105.

\bibitem{Leonidov}A.\ Leonidov, M.\ Nardi and H. Satz, \ZP C 74
(1997) 535.

\bibitem{S-W}T. {\AA}keson et al., \ZP C 46 (1990) 361.

\bibitem{Pb-Pb}N.\ Xu [NA44], \NP A 610 (1996) 175c;
P.\ G. Jones [NA49], \NP A (610) 188c.

\bibitem{Kapu-Jai}J.\ Kapusta, talk at this meeting.

\bibitem{Kapusta}S.\ Jeon and J.\ Kapusta, nucl-th/9703033,
March 1997.

\bibitem{Heinz}U.\ Heinz, \NP A 610 (1996) 264c.
\end{thebibliography}
\end{document}